\documentclass[final,5p,times,twocolumn]{elsarticle}

\usepackage{amssymb}
\usepackage[dvipdfmx]{}
\usepackage{amsmath}
\usepackage[subrefformat=parens]{subcaption}
\graphicspath{{./figures/}}

\journal{Nuclear Instruments and Methods in Physics Research Section A}

\begin{document}

\begin{frontmatter}


\author[label1]{Keigo~Yarita\corref{cor1}}
\cortext[cor1]{Corresponding author}
\ead{6217623@ed.tus.ac.jp}

\title{Proton Radiation Damage Experiment for X-Ray SOI Pixel Detectors}

\author[label1]{Takayoshi~Kohmura}
\author[label1]{Kouichi~Hagino}
\author[label1]{Taku~Kogiso}
\author[label1]{Kenji~Oono}
\author[label1]{Kousuke~Negishi} 
\author[label1]{Koki~Tamasawa}
\author[label1]{Akinori~Sasaki}
\author[label1]{Satoshi Yoshiki}
\author[label2]{Takeshi~Go~Tsuru}
\author[label2]{Takaaki~Tanaka}
\author[label2]{Hideaki~Matsumura}
\author[label2]{Katsuhiro~Tachibana}
\author[label2]{Hideki~Hayashi}
\author[label2]{Sodai~Harada}
\author[label3]{Ayaki~Takeda}
\author[label3]{Koji~Mori}
\author[label3]{Yusuke~Nishioka}
\author[label3]{Nobuaki~Takebayashi}
\author[label3]{Shoma~Yokoyama}
\author[label3]{Kohei~Fukuda}
\author[label4]{Yasuo~Arai}
\author[label4]{Toshinobu~Miyoshi}
\author[label5]{Ikuo~Kurachi}
\author[label6]{Tsuyoshi~Hamano}

\address[label1]{Department of Physics, School of Science and Technology, Tokyo University of Science, 2641 Yamazaki, Noda, Chiba 278-8510, Japan}
\address[label2]{Department of Physics, Faculty of Science, Kyoto University, Kitashirakawa-Oiwakecho, Sakyo-ku, Kyoto 606-8502, Japan}
\address[label3]{Department of Applied Physics, Faculty of Engineering, University of Miyazaki, 1-1 Gakuen-Kibanodai-Nishi, Miyazaki, Miyazaki 889-2192, Japan}
\address[label4]{Institute of Particle and Nuclear Studies, High Energy Accelerator Research Organization (KEK), 1-1 Oho, Tskuba, Ibaraki 305-0801, Japan}
\address[label5]{Department of Advanced Accelerator Technologies, High Energy Accelerator Research Organization (KEK), 1-1 Oho, Tskuba, Ibaraki 305-0801, Japan}
\address[label6]{National Institute for Radiological Sciences (NIRS), Anagawa 4-9-1, Inage-ku, Chiba-shi, Chiba, 263-8555, Japan}
\author{and the~SOIPIX~group}

\address{}

\begin{abstract}
In low earth orbit, there are many cosmic rays composed primarily of high energy protons. These cosmic rays cause surface and bulk radiation effects, resulting in degradation of detector performance. Quantitative evaluation of radiation hardness is essential in development of X-ray detectors for astronomical satellites. We performed proton irradiation experiments on newly developed X-ray detectors called XRPIX based on silicon-on-insulator technology at HIMAC in National Institute of Radiological Sciences. We irradiated 6~MeV protons with a total dose of 0.5~krad, equivalent to 6~years irradiation in orbit. As a result, the gain increases by 0.2\% and the energy resolution degrades by 0.5\%. Finally we irradiated protons up to 20~krad and found that detector performance degraded significantly at 5~krad. With 5~krad irradiation corresponding to 60 years in orbit, the gain increases by 0.7\% and the energy resolution worsens by 10\%. By decomposing into noise components, we found that the increase of the circuit noise is dominant in the degradation of the energy resolution.
\end{abstract}


\end{frontmatter}


\section{Introduction}
\label{Intro}
Radiation damage of semiconductor detectors is an important issue in high energy particle and nuclear physics experiments, in which the detectors are exposed to high radiation environments. The radiation damage is mainly due to two mechanisms: bulk effect and surface effect. The bulk effect is caused by lattice defects created by nuclear interactions between high energy particles and the detector. The defects trap carriers, and increase leakage current of the detector. The surface effect, which is often referred to as total ionizing dose (TID) effect, significantly affects MOS devices. It is due to charge accumulation in the SiO$_2$ layer induced by incident X-rays or charged particles, resulting in a shift of threshold voltages. The semiconductor detectors onboard X-ray astronomy satellites also suffer the radiation damage effect. When semiconductor detectors are used in low earth orbit at several hundred kilometers above the Earth, detector performances degrade over time. It is mainly due to cosmic-ray protons in the south atlantic anomaly (SAA) located at the altitude of $\sim 300$~km. In fact, the energy resolution of X-ray CCD onboard the Suzaku satellite degraded from 130~eV to 190~eV during the 1 year following launch~\cite{Koyama1}. Hence, quantitative evaluation of the radiation hardness is essential in development of X-ray detectors for astronomical satellites. 

\begin{figure}[tbp]
\begin{center}
\includegraphics[width=0.8\hsize]{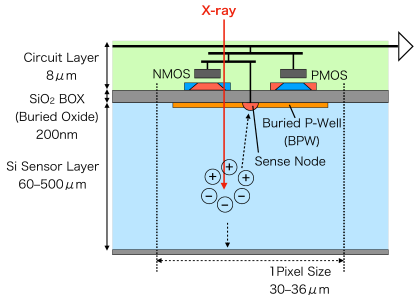}
\vspace{0cm}
\caption{Cross-sectional view of XRPIX. A thin oxide film (BOX: buried oxide) was sandwiched between sensor layer and circuit layer. This picture shows example of n-type silicon sensor layer. Holes generated by X-ray are collected to sense node. The size varies depending on series: the pixel size is 30 -- 36~$\mu$m square and the thickness of sensor layer is 60 -- 500~$\mu$m.
\vspace{0cm}
}
\label{XRPIX_image}
\end{center}
\end{figure}

We have been developing silicon on insulator (SOI) CMOS image sensor XRPIX (X-Ray PIXel)~\cite{Ryu1,Takeda1}.  Fig.~\ref{XRPIX_image} shows cross-sectional view of XRPIX. XRPIX is a monolithic detector composed of low-resistivity circuit layer, SiO$_2$ insulator layer and high-resistivity sensor layer.
X-rays are photoelectrically absorbed by the sensor layer and charges are created. These charges are collected to sense node and are converted to signals in circuit layer.
We equip the XRPIX with a function to output trigger signals and the corresponding hit-pixel addresses, with which it achieves a high time-resolution of better than $\sim $10~$\mu$s. Furthermore, the high time resolution enables the anti-coincidence method with surrounding active shields, which would effectively reduce the in-orbit background.

The main effort for development of XRPIX was poured so far into improving the energy resolution and implementing the trigger function. However in order to operate XRPIX in the space radiation environment, the effect on the performance of XRPIX by radiation damage must be understood. Therefore, a quantitative evaluation of the radiation hardness of XRPIX is necessary before the launch. Although a few studies of radiation hardness have been conducted for SOI pixel detectors \cite{Tobita1}, nothing has been done for XRPIX.

In this paper, we report on the result of the proton radiation damage experiment on XRPIX. The degradation of the energy resolution and the change of the gain are shown, and the reasons of the performance degradation are discussed.

\section{Proton radiation damage experiment for XRPIX at HIMAC }
\label{method}
We performed proton radiation damage experiment at "Heavy Ion Medical Accelerator in Chiba (HIMAC)" in National Institute of Radiological Sciences. In this experiment, we irradiated 6~MeV proton beam to XRPIX2b which was one of the XRPIX series ~\cite{Takeda1}\cite{Takeda2}. The design of XRPIX2b is summarized in Table~\ref{tab XRPIX2b}.

\begin{table}[tbp]
 \begin{center}
 \caption{The chip design of XRPIX2b used in this experiment.}
  \begin{tabular}{l c} \hline
     parameter & value\\ \hline
     Pixel size & 30~${\rm \mu}$m sq. \\
     BPW size &  12~${\rm \mu}$m sq. \\
     Type of sensor layer &  Floating zine n type Si \\
     Thickness of the sensor & 500~${\rm \mu}$m  \\
     Sensor resistivity & 5~k$\Omega \cdot$cm \\ \hline
  \end{tabular}
  \label{tab XRPIX2b}
  \end{center}
\end{table}

\subsection{Radiation dose by proton beam in this experiment}
\label{Dose}
\begin{figure}[tbp]
\begin{center}
\includegraphics[width=0.8\hsize]{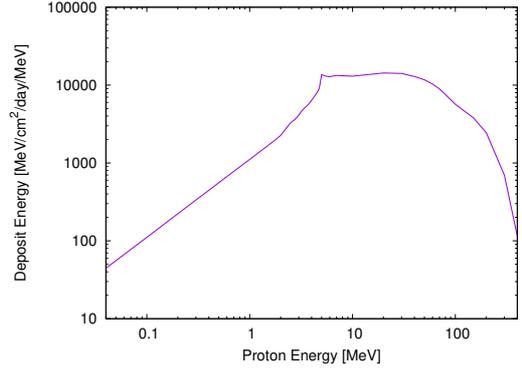}
\caption{Energy deposition in XRPIX by SAA protons. We assumed that XRPIX is surrounded by a 20 mm thick aluminum camera body, and operated in the same orbit as the Hitomi satellite.}
\label{deposit}
\end{center}
\end{figure}

In order to carry out the radiation damage experiment at the experimental facility on the ground, it is necessary to estimate the dose rate considering the space radiation environment and to convert the exposed dose in this experiment to equivalent time in space.
Firstly, we calculated the radiation dose using ESA's SPace ENVironment Information System (SPENVIS)~\cite{spenvis}, to estimate the dose rate on XRPIX assuming that XRPIX would be mounted on the future X-ray satellite in low earth orbit such as Hitomi satellite~\cite{Takahashi1}. In this calculation, we supposed the altitude and the inclination angle of the satellite were 550~km and 31$^\circ$, respectively~\cite{Mori1}\cite{Hayashi1}. We also assumed that the depletion layer thickness of XRPIX was 500~$\mu$m and XRPIX was surrounded by a 20~mm thick aluminum camera body. Secondary, we calculated the energy deposition in XRPIX by protons using stopping power value calculated by PSTAR of NIST (National Institute of Standards and Technology)~\cite{nist}.  Fig.~\ref{deposit} shows the calculated energy deposition by protons in XRPIX in orbit.  From this result, we found 4--20~MeV protons have the largest impact on XRPIX. 

Then, we calculated total proton dose by integrating the energy deposition over 0.05--400~MeV and derived the dose rate to be 1.6~$\times$~10$^6$~MeV/cm$^{2}$/day, or 82.3~rad/year. From this calculation, 0.5~krad was equivalent to 6 years operation in orbit which was the typical required lifetime for X-ray astronomical satellites. In this experiment, at first we irradiated protons up to 0.1~krad, then increased the dose little by little, and finally the total dose reached 20~krad. Eventually, we evaluated performances of XRPIX at 0.1, 0.5, 1, 5, 8, 10 and 20~krad. 
\begin{figure}[tbp]
\begin{center}
\includegraphics[width=\hsize]{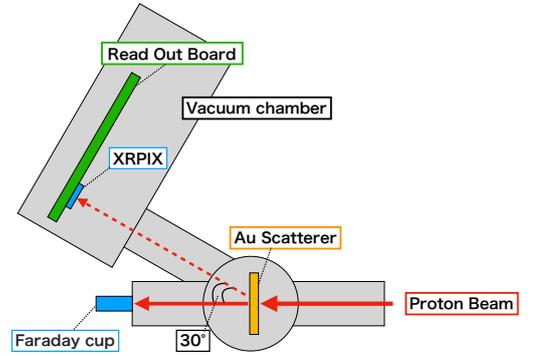}
\caption{A schematic top view of experimental set up.}
\label{ponchi}
\end{center}
\end{figure}

\subsection{Experimental setup}
\label{environment}
Fig.~\ref{ponchi} shows a schematic top view of experimental setup at the medium energy irradiation beamline
 in HIMAC. The proton beam intensity is so strong that large current induced by the protons might damage the read-out circuit on XRPIX. Therefore we set 2.5 $\mu$m thick Au film as a scatterer in front of XRPIX to reduce the proton beam intensity. XRPIX was located at 30 degree direction from the downstream and set in the vacuum chamber where the degree of vacuum was 5.0~$\times$~10$^{-4}$ hPa. From the result of simulation by Geant4~\cite{Allison1}, the proton intensity scattered to 30$^\circ$ was $6~\times~10^{-6}$ of the original beam intensity and the proton energy became $\sim$~5.9~MeV. XRPIX was cooled down to $-60^\circ$C to reproduce the radiation damage in orbit precisely.

XRPIX2b was operated with a bias voltage ($V_{\rm BB}$) of 150~V and with the integration time of 100~$\mu$s. We also evaluated the performance while changing V$_{\rm BB}$ of 5, 50, 150, 250 and 290~V. 
However, we evaluated at 8~krad only with ($V_{\rm BB}$) = 150 V.

\begin{figure}[tbp]
\begin{center}
\includegraphics[width=0.8\hsize]{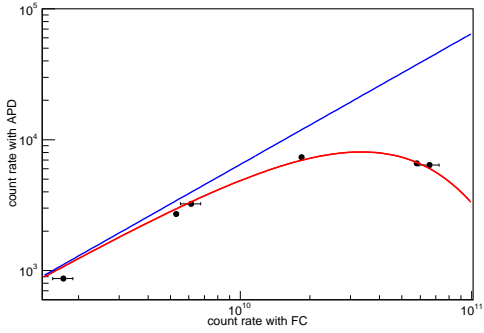}
\caption{Correlation between APD and Faraday cup. Red line is the fitting model with the paralyzed dead time model. Blue line is the model corrected dead time. The definition of bad pixels at 5~krad is shown by the blue shaded area.}
\label{APDvsFC}
\end{center}
\end{figure}

\subsection{Estimation of proton flux with APD}
Before the proton irradiation experiment for XRPIX, we irradiated protons to avalanche photodiode (APD). This APD was located at the same position as XRPIX. As shown in Fig.~\ref{ponchi}, a Faraday cup was also installed along the beam direction. It measured un-scattered protons, and tracked fluctuation of the beam intensity during the proton irradiation experiment. By comparing count rates from APD and the Faraday cup, correlation between scattered and un-scattered protons were known. 
Fig.~\ref{APDvsFC} shows the correlation of the count rate between APD and Faraday cup. The count rates from APD and the Faraday cup were not proportional because of dead time mainly due to a charge-sensitive amplifier for the APD, which behaves as a paralyzable detector. Hence, we fitted the correlation with the paralyzed dead time model expressed as
\begin{equation}
y = xe^{-x\tau},
\end{equation}
where $y$ is count rate with APD, $x$ is count rate with Faraday cup and $\tau$ is dead time~\cite{Knoll}.
The best fit model and the dead time corrected correlation between APD count rate $r_{\rm APD}$ and Faraday cup count rate $r_{\rm FC}$, $r_{\rm APD} = (6.56~\pm~0.01)~\times~10^{-7} r_{\rm FC}$ are shown in Fig.~\ref{APDvsFC}.
From this correlation and counts from the Faraday cup, we estimated the number of protons irradiated on XRPIX. The absorbed dose described hereafter is calculated based on counts from the Faraday cup.

\begin{figure}[tbp]
\begin{center}
\includegraphics[width=0.8\hsize]{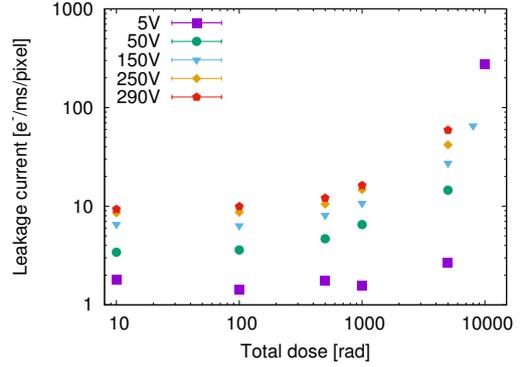}
\caption{Leakage current transition as a function of the total dose for different bias voltage (V$_{\rm BB}$). The data with V$_{\rm BB}$ of 5~V were plotted with square points. Likewise, data points were plotted for each V$_{\rm BB}$. At voltages above 50 V, leakage current increased significantly after 0.5~krad.}
\label{leak}
\end{center}
\end{figure}
\begin{figure}[tbp]
\begin{center}
\includegraphics[width=0.8\hsize]{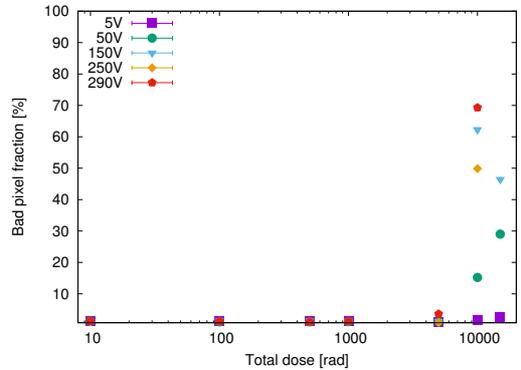}
\caption{Bad pixel fraction as a function of total dose. Almost all data points up to 1~krad, the fraction of bad pixel is about 1.3\%. At 5~krad, the fraction increased only when V$_{\rm BB}$ was 290V.  At 10~krad, most pixels were regarded as bad pixels.}
\label{badpixel}
\end{center}
\end{figure}
\begin{figure}[tbp]
\begin{center}
\includegraphics[width=0.8\hsize]{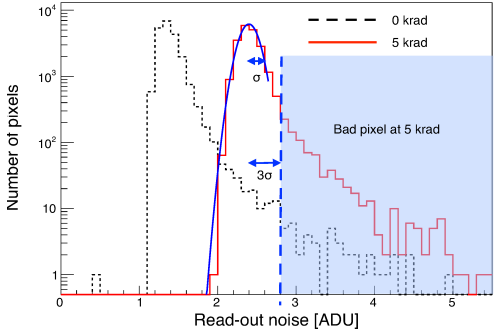}
\caption{Histogram of read-out noise for each pixel. Noise level of all pixels was increased as total dose increased. The definition of bad pixels at 5~krad is shown by V$_{\rm BB}$ shade.}
\label{readout_level}
\end{center}
\end{figure}
\begin{figure}[tbp]
\begin{center}
\includegraphics[width=0.8\hsize]{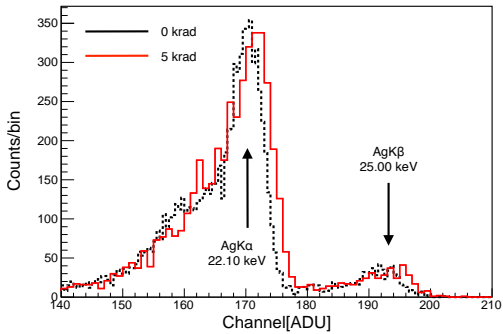}
\caption{$^{109}$Cd energy spectrum at 0~krad and 5~krad. 5~krad spectrum was scaled so that the peak has about the same counts of 0~krad peak. At 5~krad, output became higher than output at 0~krad and sigma of peak got slightly thicker.}
\label{spectrum}
\end{center}
\end{figure}

\section{Result of radiation damage experiment}
\label{result}
The leakage current was estimated by the derivative of the pedestal level with respect to the integration time, because the output signal contains the charges due to the leakage current summed during the integration time. Fig.~\ref{leak} shows the dependence of the leakage current on the total dose. Leakage current showed a tendency to increase with total dose. At 0.5~krad which is equivalent to 6~years in orbit, the leakage current increases by 1.2 times from 0~krad, and at 5~krad it rapidly increases up to 4.2 times of that at 0~krad. At 20~krad, the V$_{\rm BB}$ was applied up to only 5~V because the ADC output were saturated above 5~V.

\begin{figure*}[tbp]
\begin{minipage}{0.5\hsize}
\begin{center}
\includegraphics[width=0.8\hsize]{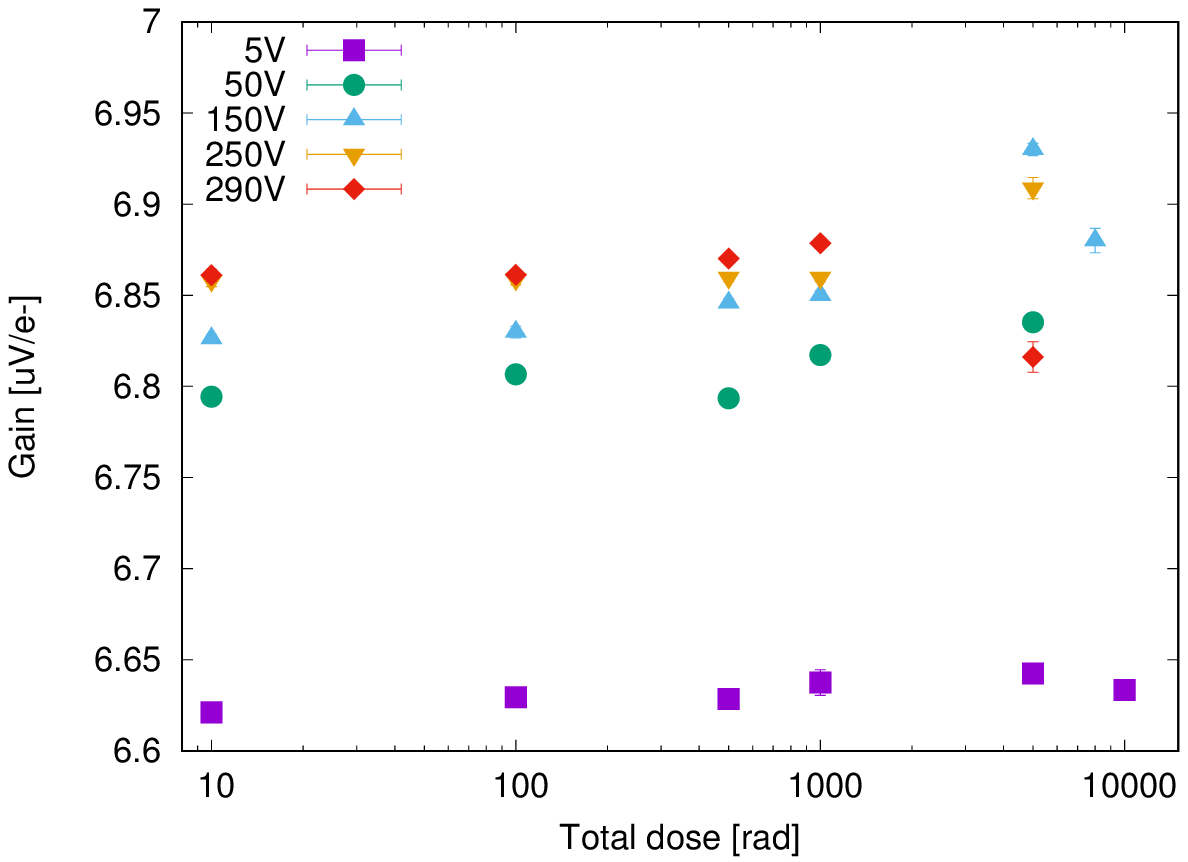}
\caption{Gain transition as a function of total dose. Gains obtained by $^{109}$Cd AgK$\alpha$ peaks. The gain increased significantly at 5~krad. Furthermore the gain decreased at 5~krad with V$_{\rm BB}$ 290V and after 8~krad.}
\label{gain}
\end{center}
\end{minipage}
\begin{minipage}{0.5\hsize}
\begin{center}
\includegraphics[width=0.8\hsize]{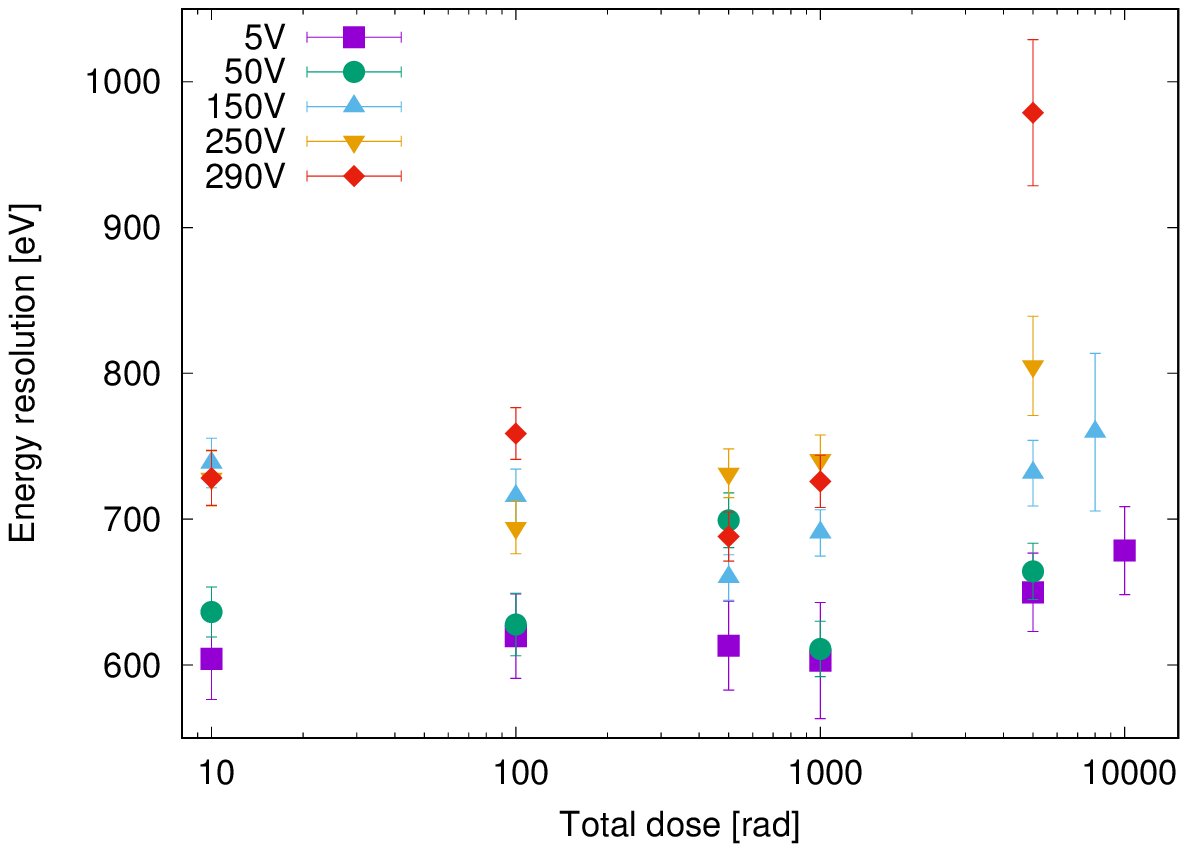}
\caption{Energy resolution transition as a function of total dose. Energy resolution increased significantly at 5~krad.}
\label{fwhm}
\end{center}
\end{minipage}
\end{figure*}

The dependence of bad pixel fractions on total dose is shown in Fig.~\ref{badpixel}. 
The bad pixel is defined by using distribution of read out noise as shown in Fig.~\ref{readout_level}. Standard deviation $\sigma$ of the distribution was evaluated by fitting the distribution with Gaussian function. Then, pixels outside of $3\sigma$ are defined as the bad pixel. They are eliminated from all the spectral analyses described below. The number of bad pixels increased rapidly at 10~krad because variations in readout noise between pixels increased. By using this definition, most pixels were regarded as bad pixels.

Fig.~\ref{spectrum} shows the energy spectra obtained by $^{109}$Cd radioactive source at 0~krad and 5~krad. In extraction of the spectra, the data were analyzed in the same manner as described in~\cite{Nakashima1}. Pulse heights exceeding the event threshold were extracted, and then those from surrounding 8 pixels exceeding the split threshold are merged. In this result, spectra of single pixel events, in which all the pulse heights from surrounding 8 pixels were below the threshold, were shown. 
Since the Ag~K$\alpha$ line at 22.10~keV is much stronger than the Ag~K$\beta$ line in $^{109}$Cd spectra, only K$\alpha$ line was used in the spectral analysis. So that only AgK$\alpha$ peak was used when we evaluated performance.
The Ag~K$\alpha$ line consists of Ag K$\alpha_1$ and Ag K$\alpha_2$, whose energy separation is as large as $\simeq170$~keV. If the spectra were fitted with single Gaussian model, the energy resolution would be overestimated. Thus, we evaluated the gain and energy resolution by fitting with 2 Gaussians in which sigma is tied each other and mean is linked with energy ratio. 2 Gaussians expressed as
\begin{equation}
\begin{split}
f(x) = &~AI_{\rm K\alpha_{1}}\exp\left(-\frac{(x-GE_{\rm K\alpha_{1}})^2}{2\sigma^2}\right)\\
& +AI_{\rm K\alpha_{2}}\exp\left(-\frac{(x-GE_{\rm K\alpha_{2}})^2}{2\sigma^2}\right)
\end{split}
\end{equation}
where $I_{\rm K\alpha_{1,2}}$ and $E_{\rm K\alpha_{1,2}}$ are the emission probability and line energy of Ag K$\alpha_{1,2}$ line. Normalization $A$, width $\sigma$ and gain $G$ were free parameters.
Fig.~\ref{gain} shows the dependence of gain on total dose. The datapoint at 10~rad represents the value before proton irradiation. With 1~krad irradiation or less, the gain at V$_{\rm BB}$ = 250~V was consistent with that before the irradiation. At 5~krad, the gain increased by 0.7$\pm$0.1\%. Furthermore, the gain decreased when we applied high back bias voltages after 5~krad irradiation.

We show transition of energy resolution in Fig.~\ref{fwhm}. After 5~krad, error bars became larger due to the increase of the number of bad pixels. As with the gain, the energy resolution increased at 5~krad by 10.6$\pm$5.8\%.
From these results, though the leakage current increased to 1.2 times at 0.5~krad, detector performance such as gain and energy resolution did not change, indicating that XRPIX is tolerant for 6 year operation in space. 

\section{Discussion}
\label{discussion}

\subsection{The influence of bulk effect and surface effect}
\label{The cause of radiation damage}
As described in section~\ref{Intro}, the detector performance degrades with proton irradiation via two different processes, namely surface and bulk effects. Whereas the surface effect strongly depends on the detector configuration, the bulk effect is roughly determined by only the particle fluence. Thus, we estimated the influence of the bulk effect by protons in orbit. According to~\cite{Ohsugi1}, by the bulk effect, an increase of leakage current per unit volume $\Delta I$ at a temperature of $25^\circ$C is written as
\begin{equation}
I_{\rm increase} = 3.0~\times~10^{-8}~\times~F~{\rm nA/cm}^3{\rm /year},
\end{equation}
where $F$ in this expression is proton fluence. 
From our estimation described in section~\ref{Dose}, the proton fluence is expected to be $F = 4.9~\times~10^7${\rm ~/cm}$^{2}$/year in orbit, then the increase of the leakage current is  $\Delta_I$ = 6.6$~\times~$10$^{-16}$~A/year/pixel at 25$^\circ$C, corresponding to ~5.9$~\times~10^{-20}$ A/pixel/year at $-60^\circ$C on the assumption of the typical temperature dependence $I~\propto~\exp(-E_{\rm g}/2kT)$. 
In our experiment, leakage current increased from $1.4~\times~10^{-15}$ A/pixel to $1.7~\times~10^{-15}$ A with 0.5~krad irradiation. Clearly, it is much higher than those expected by the bulk effect, indicating that the surface effect is dominant in XRPIX.

Since the surface effect is dominant in the radiation damage of XRPIX by the proton irradiation, the total dose on the insulator layer is more important than those on the sensor layer estimated in section~\ref{Dose}.

Therefore, we calculated energy deposition in 0.2 $\mu$m SiO2 insulator layer. As a result, the energy deposition on the insulator layer in orbit is estimated as 4.1$~\times~10^{-2}$ rad/year, which is very close to that on the sensor layer. In addition, 6 MeV protons used in our experiment deposit 9.42$~\times~10^{-11}$~rad on the sensor layer and 4.57$~\times~10^{-14}$~rad on the insulator layer. Therefore, in this experiment, the radiation damage for the insulator layer is consistent to that for the sensor layer within 6.5\%.

\subsection{Decomposition into noise components}
\label{noise_estimation}
To identify the origin of the energy resolution degradation, we decomposed into noise components by using the following expression~\cite{Knoll}\cite{Sze}.
\begin{equation}
  \sigma^2 = \sigma_{\rm Fano}^2 + \sigma_{\rm RO}^2 .
\end{equation}
$\sigma_{\rm Fano}$ is Fano noise, and $\sigma_{\rm RO}$ is read-out noise. The read-out noise consists of CMOS circuit noise $\sigma_{\rm cir}$ and shot noise of leakage current $\sigma_{\rm leak}$,
\begin{equation}
  \sigma_{\rm RO}^2 = \sigma_{\rm cir}^2 + \sigma_{\rm leak}^2.
\end{equation}
Thus, the energy resolution is expressed as
\begin{equation}
  \sigma^2  = \sigma_{\rm Fano}^2 + \sigma_{\rm cir}^2 + \sigma_{\rm leak}^2  + \sigma_{\rm other}^2,
\end{equation}
where the last term $\sigma_{\rm other}^2$ is introduced to reproduce the energy resolution.

\begin{figure}[tbp]
\begin{center}
        \includegraphics[width=0.9\hsize]{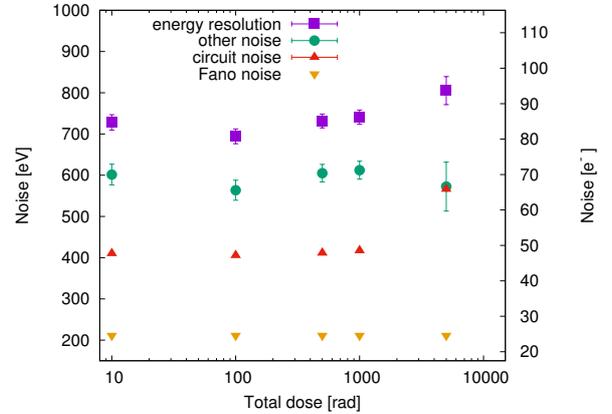}
  \caption{Noise components of energy resolution with V$_{\rm BB}$ = 250~V. Fano factor was regarded as 0.1.}
  \label{noise}
  \end{center}
\end{figure}

By using this expression, we decomposed the energy resolutions into each noise component as shown in Fig.~\ref{noise}.
The shot noise of the leakage current is not shown in Fig.~\ref{noise} because it is so small as $~\sim~1$~eV. From this result, only circuit noise increases as the total dose increases. In other words, degradation of energy resolution were caused by increase of circuit noise.

\section{Summary}
\label{conclusion}
We irradiated 6~MeV proton beam to XRPIX2b. The detectors performance was almost unchanged at 0.5~krad which is equivalent to 6 years in orbit. At 5~krad, the energy resolution was increased significantly and the leakage current was increased by 5 times from 0~krad, but increase of gain was only 0.7\%. By decomposing the energy resolution into noise components, we found that CMOS circuit increased the noise.

\section*{Acknowledgements}
\label{Acknowledgements}
We acknowledge the valuable advice and great work by the personnel of LAPIS Semiconductor Co., Ltd. This study was supported by the Japan Society for the Promotion of Science (JSPS) KAKENHI Grant-in-Aid for Scientific Research on Innovative Areas 25109002 (Y.A.), 25109003 (S.K.), 25109004 (T.G.T. and T.T.), 20365505 (T.K), 23740199 (T.K), 18740110 (T.K.), Grant-in-Aid for Young Scientists (B) 15K17648 (A.T.), Grant-in-Aid for Challenging Exploratory Research 26610047 (T.G.T.) and Grant-in-Aid for JSPS Fellows 15J01842 (H.M.). This study was also supported by the VLSI Design and Education Center (VDEC), the University of Tokyo in collaboration with Cadence Design Systems, Inc., and Mentor Graphics, Inc.




\begin{thebibliography}{20}

\bibitem{Koyama1}
K. Koyama et al., X-Ray Imaging Spectrometer (XIS) on Board Suzaku, Publications of the Astronomical Society of Japan 59 (2007) S23.
\bibitem{Ryu1}
S. G. Ryu et al., First Performance Evaluation of an X-Ray SOI Pixel Sensor for Imaging Spectroscopy and Intra-Pixel Trigger, IEEE Transaction on Nuclear Science 58 (2011) 2528.
\bibitem{Takeda1}
A. Takeda et al., Development and Evaluation of an Event-Driven SOI Pixel Detector for X-Ray Astronomy, in proceedings of Technology and Instrumentation in Particle Physics 2014, June, 2-6, 2014 Amsterdam, the Netherlands PoS(TIPP2014)138.
\bibitem{Arai1}
Y. Arai et al., Development of SOI pixel process technology, Nuclear Instruments and Methods in Physics Research Section A 636 (2011) S31.
\bibitem{Tobita1}
N. Tobita et al., Compensation for TID Damage in SOI Pixel Devices, Proceedings of International Workshop on SOI Pixel Detector (SOIPIX2015), Tohoku University, Sendai, Japan, 3-6, June, 2015.
\bibitem{Takeda2}
A. Takeda et al., Design and Evaluation of an SOI Pixel Sensor for Trigger-Driven X-ray Readout, IEEE Transaction on Nuclear Science 60 (2013) 586.
\bibitem{spenvis}
SPENVIS, The Space Environment Information System, https://www.spenvis.oma.be/regulation.php
\bibitem{Takahashi1}
T. Takahashi et al., The ASTRO-H (Hitomi) x-ray astronomy satellite, Proc. SPIE, 99050U (2016)
\bibitem{Mori1}
K. Mori et al., Proton Radiation Damage Experiment on P-Channel CCD for an X-ray CCD camera onboard the Astro-H satellite, Nuclear Instruments and Methods in Physics Research Section A 731 (2013) 160
\bibitem{Hayashi1}
K. Hayashi et al., Radiation effects on the silicon semiconductor detectors for the ASTRO\-H mission, Nuclear Instruments and Methods in Physics Research A 699 (2013) 225\-229
\bibitem{nist}
PSTAR,stopping-power and range tables for protons,https://physics.nist.gov/PhysRefData/Star/Text/PSTAR.html
\bibitem{Allison1}
J. Allison et al., Geant4 Developments and Applications, IEEE TRANSACTIONS ON NUCLEAR SCIENCE, VOL. 53, NO. 1, FEBRUARY 2006
\bibitem{Nakashima1}
S.Nakashima et al., Progress in Development of Monolithic Active Pixel Detector for X-ray Astronomy with SOI CMOS Technology,  Physics Procedia 37 (2012) 1392.
\bibitem{Barnaby1}
H. J. Barnaby et al., Total-Ionizing-Dose Effects in Modern CMOS Technologies, IEEE TRANSACTIONS ON NUCLEAR SCIENCE, VOL. 53, NO. 6, December 2006
\bibitem{Ohsugi1}
T.Ohsugi et al., Radiation damage in silicon microstrip detectors, Nuclear Instruments and Methods in Physics Research Section A 265 (1988) 105\-111
\bibitem{Knoll}
Glenn F. Knoll, RADIATION DETECTION AND MEASUREMENT, 4th Edtion, John Wiley $\&$ Sons, Inc. (2010)
\bibitem{Sze}
S. M. Sze, SEMICONDUCTOR DEVICES, 2nd Edition, John Wiley $\&$ Sons, Inc. (2002)

\end{thebibliography}

\section*{References}
\label{References}

\end{document}